\documentclass[pre,twocolumn,showpacs,amsfonts,amsmath,preprintnumbers,amssymb,superscriptaddress]{revtex4}

\usepackage{graphicx}
\usepackage{dcolumn}
\usepackage{bm}

\graphicspath{{figures/},.}

\begin{document}

\preprint{APS/123-QED}

\title{Synchronization in symmetric bipolar population networks}

\author{Lubos Buzna}
\email{buzna@frdsa.uniza.sk}
\affiliation{ETH Zurich, UNO C 14, Universit\"atstrasse 41, 8092 Zurich, Switzerland}
\affiliation{University of Zilina, Univerzitna 8215/5, 01026 Zilina, Slovakia}

\author{Sergi Lozano}
\affiliation{ETH Zurich, UNO C 14, Universit\"atstrasse 41, 8092 Zurich, Switzerland}

\author{Albert D\'{\i}az-Guilera}
\affiliation{ETH Zurich, UNO C 14, Universit\"atstrasse 41, 8092 Zurich, Switzerland}
\affiliation{Potsdam Institute for Climate Impact Research (PIK), Telegraphenberg A 31, 14473 Potsdam, Germany}
\affiliation{Center for Dynamics of Complex Systems (DYCOS), Potsdam, Germany}
\altaffiliation{\emph{Permanent address:}  Departament de F\'{\i}sica Fonamental, Universitat de Barcelona, 08028 Barcelona, Spain}
\begin{abstract}
We analyze populations of Kuramoto oscillators with a particular distribution of natural frequencies. Inspired by networks where there are two groups of nodes with opposite behaviors, as for instance in power-grids where energy is either generated or consumed at different locations, we assume that the frequencies can take only two different values. Correlations between the value of the frequency of a given node and its topological localization are considered in both regular and random topologies. Synchronization is enhanced when nodes are surrounded by nodes of the opposite frequency. 
The new theoretical result presented in this paper is an analytical estimation for the minimum value of the coupling strength between oscillators that guarantees the achievement of a globally synchronized state. This analytical estimation, which is in a very good agreement with numerical simulations, provides a better understanding of the effect of topological localization of natural frequencies on synchronization dynamics. 
\end{abstract}
\maketitle
%
\section{Introduction}
\label{sec1}
During the last decade, the realization that many observed complex systems in nature and society are intrinsically related to nontrivial patterns of interactions has created a great interest in what we nowadays know as complex networks theory \cite{s01,ab02,dm02,n03a,blmch06}. Once the basic statistical properties of the individual constituents (degree distributions, clustering coefficients, distances, diameters and so on) were analyzed in detail (see Ref. \cite{crtbv07} for a survey of measurements) the attention turned to the correlations among neighbors. Examples of this are degree assortativity \cite{n02}, degree-degree correlations \cite{sbpv07} or, even, what is known as the rich-club coefficient \cite{rich-club}. From the topological point of view, at higher scales complex networks are usually organized in groups that have a  dense internal connectivity. These groups are known as functional groups in biology or as network communities in the social sciences. The analysis of these mesoscopic structures has also received much attention from the statistical physics community \cite{gn02,ddda05}.
It has been also clarified how topological complexity gives rise to a complex dynamical behavior. Complexity results not only from the appearance of emergent behaviors from simple dynamical units, but an important part of this collective behavior lies precisely in the interaction patterns \cite{blmch06,indiana_book}. Among the many dynamical behaviors that have been considered in the statistical physics literature, one of the most intensively analyzed is that of synchronization. We should understand the phenomena of synchronization as an emergent cooperative behavior, where units with similar individual behaviors develop a coordinated collective output, in which all units follow the same evolution in time \cite{strogatz03}. One of the paradigmatic models for the explanation of this widely occurring phenomenon goes back to Kuramoto \cite{Kuramoto84,abprs05}. In his model, units are characterized only by their phases $\varphi_i$, which evolve according to the equation
\begin{equation}
\label{kuramoto}
\dot{\varphi}_i=\omega_i + \sigma \sum_j a_{ij} \sin (\varphi_j - \varphi_i).
\end{equation}
Here $\omega_i$ is the natural frequency of unit $i$, $\sigma$ is the coupling strength, and $a_{ij}$ is the connectivity matrix ($a_{ij}=1$ if $i$ and $j$ are connected, 0 otherwise). This set of equations gives rise to two qualitative behaviors that can be easily understood from the model. The first term makes the units to follow their natural frequencies $\omega_i$ (in general these frequencies are not identical, but given by a certain distribution). The second one makes the phases to approach each other. If the first term dominates we have incoherent behavior given by a distribution of effective frequencies, $\dot{\varphi_i}$, whose precise values depend strongly on the network topology.

If the second one dominates, we observe a coherent behavior where all the phases approach, and the effective frequencies become equal, settling the system into a synchronized state. Consequently, there must be a transition from an incoherent to a synchronized state. This transition is typically characterized by an order parameter $r$, defined through the equation
\begin{equation}
\label{order_parameter}
r e^{i\Psi}=\sum_j e^{i\varphi_j},
\end{equation}
where $\Psi$ is a global phase (not constant) \cite{Kuramoto84,abprs05}.
This transition has been widely analyzed for symmetric distributions of frequencies for regular \cite{abprs05} as well as complex \cite{adkmz08} topologies. Frequencies are usually assumed to be
randomly distributed across the network, and no particular consideration about the correlation between frequency and network location has been made until quite recently. In particular, in Ref. \cite{brede07}  a uniform distribution of natural frequencies in different topologies has been considered.  It has been found that, when frequencies between neighbors are negatively correlated (i.e. nodes with positive frequency, taken from a symmetric interval, tend to be surrounded by nodes with negative frequencies) the network is more easily synchronized. Moreover, Ref.  \cite{gz06} studied the synchronization in Erd\"os-R\'enyi networks when adaptively changing the topology of the interaction network to enhance synchronization. One observes the emergence of clusters, characterized by similar  values of nominal frequencies,  which were able to synchronize individually significantly below the global synchronization threshold. The resulting interaction network eventually changed its topology. Namely it evolved from its initial random structure towards a small-world pattern.

Following this research line, we will analyze here the correlation (or anticorrelation) between frequencies and localization in the network, but in a somewhat different framework. We assume that the frequencies of the oscillators are distributed in a bipolar way, having values either -1 or +1. We take +1 and -1 without loss of generality. On the one hand, the fact that the distribution of frequencies is bipolar does not imply it to be symmetric in general. Nevertheless the frequency distribution can be shifted arbitrarily by transforming it into a comoving frame of reference and the interest lies only on the phase and frequency differences and not on absolute values. On the other hand, we are only interested on the relative importance of the two terms, the frequency values and the coupling strength; fixing one of these values (+1 and -1 in our case) just fixes the time scale. This particular distribution (which is mathematically described as the sum of two delta functions) can also be understood as a limiting case of a bimodal distribution of natural frequencies, which has been analyzed so far in populations of all-to-all coupled Kuramoto oscillators \cite{Martens,Montbrio,bps98}, i.e. without considering network effects.

Also from a practical point of view, this bipolar distribution can be related to the electrical power-grid, where energy is either generated or consumed and hence nodes play complementary roles. Actually, generators have natural frequencies slightly above the network frequency (50 or 60 Hz) while machines have natural frequencies slightly below. If we consider the system in a co-rotating frame (at 50 Hz) it turns out that generators have positive small frequencies and machines have negative small frequencies, that corresponds to the bipolar distribution considered in our work. As it is shown in Ref. \cite{fnp07}, Kuramoto-like equations can be derived from swing and power flow equations using various simplications, as for example, assuming uniform maschines and generators. Although, a new ingredient (an inertia term) enters the description, which can be of crucial importance \cite{Acebron,Tanaka}. The final goal of the powergrid is to work in a globally synchronized way and the lack of synchronization can result in serious damages.

Finally, as we will show later, a bipolar distribution allows one to calculate analytically the synchronization threshold in a regular lattice and to determine it approximately in the general case. This last point is, actually, a key contribution to the literature on dynamics on networks, where the number of exact results is scarce.
Our paper is organized as follows: Section \ref{sec2} introduces the dynamical model and the procedure used to distribute the natural frequencies over the network. Then, the influence of the natural frequencies distribution on two models of random networks (a regular random network and a Barabasi-Albert scale-free network) is analyzed numerically in Sect. \ref{sec3}. This study continues in Sect. \ref{sec4} by carrying out analytical calculations for the case of regular networks and is extended in Sect. \ref{sec5}, providing a good approximation for the case of random networks. Section \ref{sec6} summarizes our conclusions.

\section{Dynamical model and reshuffling procedure}
\label{sec2}
We consider an undirected graph ${\cal G}$ which is composed of a set of $N$ nodes and a set of $L$ links, and whose adjacency matrix is $a_{ij}$. At each node of the network there is an oscillator whose phase evolves according to Eq. (\ref{kuramoto}) and is, hence, coupled to a set of neighbors. We assign natural frequencies +1 to half of the population and -1 to the other half at random. Starting from initial random phases, we let the system evolve for a time long enough to reach a stable state in which all oscillators have a bounded frequency $\dot{\varphi}_i$ (which in general is not constant).

Since the in the interaction term of  the oscillators equations is an odd function and the connectivity matrix is symmetric, summing up the complete set of equations of motion we always get
\begin{equation}
\sum_i\dot{\varphi}_i=\sum_i \omega_i = 0.
\end{equation}
Therefore, one possible solution is $\dot{\varphi}_i=0$ for all  $i$, which corresponds to the synchronized state. However, in order for this state to be reached, the coupling needs to be strong enough to overcome the natural frequency distribution (in our case either +1 or -1). Hence, for weak coupling we expect the system to develop a nonhomogeneous distribution of effective frequencies $\dot{\varphi}_i$, corresponding to a non-synchronized state.
A transition from one state to the other is produced at some critical value of the coupling strength, whose determination is one of the goals of this paper. In order to check, in a controllable way, the dependence of the critical value on the precise location of the respective oscillators, we introduce a parameter that quantifies this correlation. For a single node, we define its {\em frequency similarity} as
\begin{equation}
\label{node_similarity}
 S_i=\frac{N_i(\omega_i)}{N_i},
\end{equation}
where $N_i$ is the number of neighbors of node $i$ and $N_i(\omega_i)$ specifies how many out of them have the same natural frequency as node $i$. Furthermore, we define the overall  {\it frequency similarity of the network} by the mean value
\begin{equation}
\label{network_similarity}
 S_{\cal G} = \langle S_i \rangle.
\end{equation}
In a similar framework, Brede \cite{brede07} analyzed a network of Kuramoto oscillators in which the natural frequencies were taken from a uniform distribution in the interval $[-1,+1]$. By switching the frequencies of neighboring oscillators such that they tend to be anticorrelated, the author showed that this procedure enhances the synchronizability of the networks in the sense that the critical coupling necessary for the synchronization transition is decreased.

\begin{figure}
\includegraphics[width=0.45\textwidth]{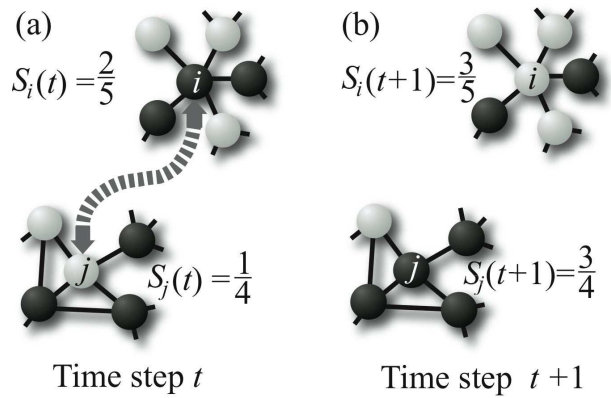}
\caption{ Illustration of the swap operation between nodes $i$ and $j$ in the reshuffling procedure. To represent visually the bimodal distribution of frequencies, we assign colors to nodes according to their nominal frequencies $\omega_i$. Column (a) corresponds to the iteration step $t$; two nodes $i$ and $j$ are chosen at random and we compute their frequency similarity $S_i$ and $S_j$. Column (b) represent the iteration step $t+1$, where the frequencies of the nodes have been interchanged since the overall network similarity increases.}
\label{fig:reshuffling}
\end{figure}

In the following we will propose a mechanism that does both, either maximizes or minimizes the frequency similarity of the network and hence allows us to  study its impact on the network dynamics. We proceed as follows \footnote{The procedure can be enhanced, for instance, by applying a simulated annealing approach, where the rule deciding whether to accept or to reject the swap operation is probabilistic. Moreover, the initialization can speed up the maximization or minimization of network similarity when initial positions for each color (i.e. for each natural frequency) are chosen. For example, "flooding" the network with fitting colors starting from a selected node, gives already close-to-optimal configurations.}:
\begin{enumerate}
\item
First, we randomly assign the frequency $+1$ to one half of nodes and the frequency $-1$ to the other half.  Then, we measure the value of network similarity.
\item
If either the current value of similarity is close enough to the desired value or the number of iterations exceeded a predefined limit, the procedure is terminated. Otherwise we continue to the next step.
\item
We randomly choose two nodes $i$ and $j$ with different frequencies and we evaluate whether the exchange of their frequencies increases or decreases the value of network similarity (see Fig. \ref{fig:reshuffling}). When the new value of similarity is closer to the desired one, the swap operation is accepted and the algorithm continues with the processing of step 2.
\end{enumerate}

\section{Simulation results  for random networks}
\label{sec3}
Following the procedure outlined above, we have performed simulations for two types of random topologies. On the one hand scale-free (SF) networks grown by the preferential attachment mechanism proposed by Barabasi and Albert \cite{ba99}. On the other hand random regular (RR) graphs for which the number of links of each node is constant but randomly distributed to the rest of nodes \cite{kim_2003}. In both cases we take $N=500$ nodes and an average connectivity equal to 3 (for the RR graph, 3 is the number of links for each node).
\begin{figure}
\includegraphics[width=0.4\textwidth]{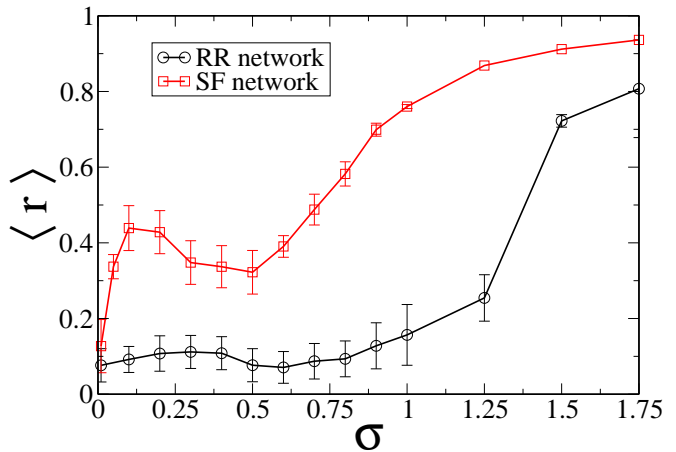}
\caption{
\label{fig:Fig_r_rand} 
Order parameter $r$ (see Eq. \ref{order_parameter}) as a function of the coupling strength $\sigma$  for random allocations of natural frequencies on two types of networks. Black and red solid lines correspond, respectively, to a Random Regular (RR) network with similarity $S_{\cal G}=0.44$ and a Barabasi-Albert scale-free (SF) network with similarity $S_{\cal G}=0.624$ (Number of nodes $N = 500$ in both cases). Each point is an  average over 30 independent simulations with initially uniformly distributed random phases, and the error bars show the standard deviation.}
\end{figure}

In Fig. \ref{fig:Fig_r_rand}  we show the usual order parameter $r$ as a function of coupling strength $\sigma$, Eq. (2), for these two types of networks. Unexpectedly,  $r$ values do not increase monotonically with $\sigma$, but show a local maximum at low values of $\sigma$ (around 0.25 for the RR and 0.1 for the SF). This phenomenon can be explained as a result of local synchronization among neighboring oscillators with the same natural frequency when $\sigma$ is small. As the coupling strength is high enough to let these oscillators to interact, they tend to synchronize their phases, leading to higher values of the $r$ order parameter (which is a measure of global phase similarity). This effect disappears for higher values of $\sigma$, as the system evolves towards a global synchronization.  Consequently, taking into account that $r$ does not clearly differentiate between local and global synchronization,  we adopt the following order parameter:

\begin{equation}
\label{eq:calculation_sigma}
r_\omega = \sqrt{\frac{1}{N}\sum_{i \in N}[\dot{\varphi}_i-\langle\omega\rangle]^2},
\end{equation}
which is a measure of the effective frequency dispersion  \cite{Nadal}. In our case, for a coupling strength of zero the order parameter is 1, whereas in the synchronized state it is equal to 0, facilitating a clear distinction between these two states.

\begin{figure}
\includegraphics[width=0.5\textwidth]{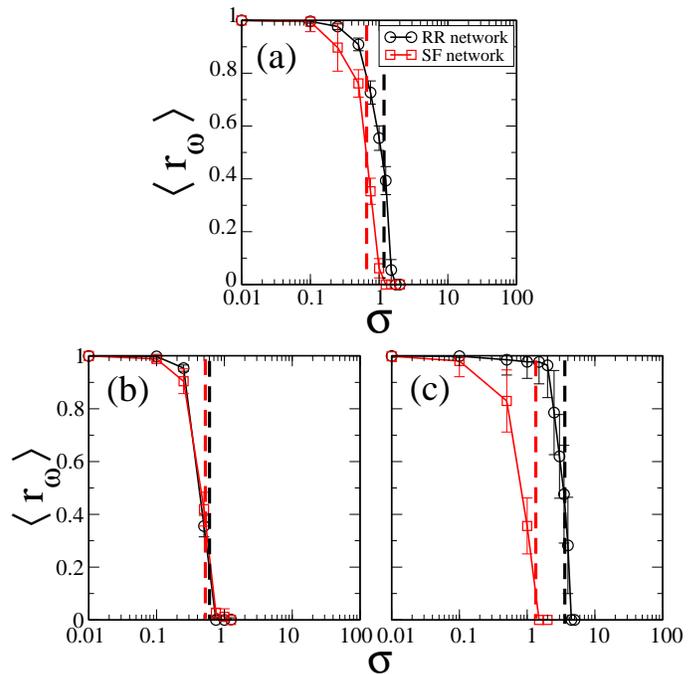}
\caption{Synchronization (averaged stationary value of the proposed order parameter $r_\omega$) as a function of the coupling strength $\sigma$, for RR and SF networks ($N = 500$ and $\bar{k}=3$) and different network similarities. (a) original network similarity ($S_{RR} = 0.467 , S_{SF}  = 0.475$). (b) low network similarity ($S_{RR}=0.099, S_{SF}  =  0.144$).  (c) high network similarity($S_{RR} = 0.904, S_{SF}  = 0.859$). Each point corresponds to an average over 30 independent simulations with uniformly distributed random phases as initial conditions and the error bars stand for the standard deviation. The vertical dashed lines correspond to the approximate analytical solutions for the critical value as calculated in Sec. \ref{sec5}.}
\label{fig:exp_RR+SF}
\end{figure}

In the three panels of Fig.\ref{fig:exp_RR+SF}, we can see how the transition can be very well characterized by this new order parameter. In the top panel we show the network with the random allocation of the natural frequencies, whereas in the lower panels we have the behaviors corresponding to the extreme values of network similarity. From this comparison, we get a two-fold message. First, SF networks synchronize more easily than RR networks, which is in agreement with the results of Ref. \cite{gma07a}. Second, low similarity enhances synchronizability whereas high similarity makes it harder. This last conclusion is the same that was obtained in Ref. \cite{brede07} when dealing with global synchronization. The same author, however, also realized that high similarity can be better for local synchronization \cite{brede08}. The phenomena of local synchronization was also analyzed in a temporal perspective in Ref. \cite{adp06a}. In this paper, a population of identical oscillators gets partially synchronized at different time scales that can be related to topological scales. In the case presented here of nonidentical oscillators, this could be achieved by introducing a discrete (but large) set of natural frequencies, one for each cluster to be synchronized. Similar results (i.e., formation of cluster with similar frequencies) were also observed in Ref. \cite{gz06}, as the links were locally reshuffled, increasing the  similarity of frequencies in the local neighborhood of nodes.

Here we notice that for low similarity the two network types show almost the same behavior, while for high similarity, which is in general worse for the synchronization, the SF network synchronizes more easily than the RR one. Actually, the SF network shows only a small effect of the frequency similarity of the network but for the RR network it is evident. These results can be explained from synchronization among neighboring oscillators. When the frequency similarity of the network is small the system is more easily  synchronized because the interaction is more effective, phases of neighboring units can be larger. On the contrary, for a large frequency similarity nodes are surrounded by nodes of the same frequency and the system is locally stable; however this local synchronization is responsible for the difficulty in achieving the global one. Oscillators only interact along the borders of regions of opposite frequencies and the core of these regions burdens the advance of the synchronization wave, then  the larger the regions of same frequency the harder to be synchronized.

\section{Analytical approach for regular lattices}
\label{sec4}
\begin{figure}
\includegraphics[width=0.5\textwidth]{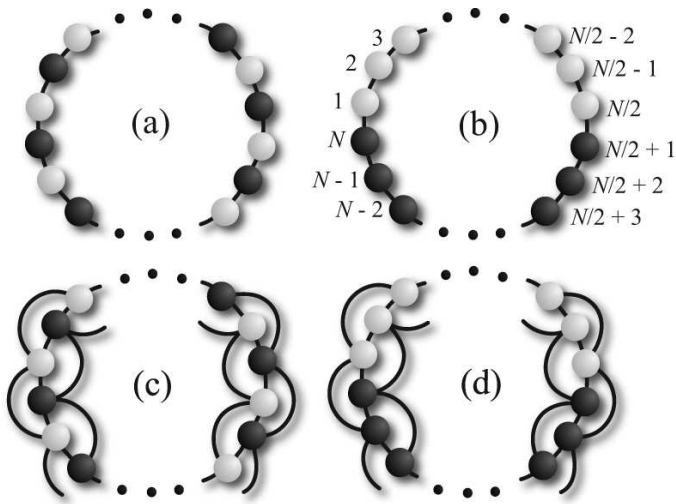}
\caption{Distributions of natural frequencies with low (left) and high (right) network similarity on two regular lattices with node degrees 2 (top) and 4 (bottom). White and black colors correspond, respectively, to $+1$ and $-1$ natural frequencies. See main text for details.}
\label{fig:regular_lattices}
\end{figure}
Analyzing the above mentioned effects in regular one-dimensional lattices has several advantages:
i) we can construct ad-hoc networks with extreme values of the network similarity parameter, ii) we can perform
some exact calculations, and iii) we can test how well the exact results serve as an approximation for general random networks.

Let us first consider a 1-dimensional ring  with only nearest neighbor connections. The two extreme cases are:\\

\noindent a) a ring with alternating natural frequencies +1 and -1 (see Fig. \ref{fig:regular_lattices}(a)) and in this case $S_{\cal G}=0 $.
For symmetry reasons, we can assume that all units with natural frequency +1 have the same frequency $\dot{\varphi}_+$ and phase
${\varphi}_+$ in the synchronized stationary state. The same holds for the negative ones with
frequency $\dot{\varphi}_-$ and phase
${\varphi}_-$.
Then it is easy to find the following relation
for any node with positive frequency:
\begin{equation}
\label{1d_0sim}
\dot{\varphi}_+= 1 + 2\sigma \sin (\varphi_- - \varphi_+).
\end{equation}
In the synchronized state the left-hand-side of this equation is 0.
Then the coupling has to verify
\begin{equation}
\label{1d_0simo}
\sigma = \frac{1}{2 \sin (\varphi_+ - \varphi_-)}.
\end{equation}
and since the sine function is bounded $\sigma > 1/2$, defining the critical value of $\sigma$, $\sigma_c=1/2$,
above which synchronization can exist (it is a necessary but not sufficient condition).

\noindent b) In this case we have half the ring with positive natural frequencies and the other half with
negative natural frequencies, and hence $S_{\cal G}=1-2/N $. We number the positive natural frequencies in clockwise direction starting from the leftmost point (white-colored in Fig. \ref{fig:regular_lattices}(b))
Summing up the set of equations
(1) from 1 to $N/2$ all coupling terms cancel each other out apart from the first and the last ones:

\begin{equation}
\label{1d_1sim}
\dot{\varphi}_1 + \cdots + \dot{\varphi}_{\frac{N}{2}}=
 \frac{N}{2} + \sigma \sin ({\varphi}_{\frac{N}{2}+1} - {\varphi}_{\frac{N}{2}}) + \sigma \sin (\varphi_{N} - \varphi_1).
\end{equation}

Again, if the system is synchronized, the left-hand side of the equation equals 0. On the right-hand side, we find two equivalent contributions to the coupling, the two links connecting oscillators sitting on the border of the two regions (defined by the two different values of natural frequencies). We find that the largest phase difference between any pair of nodes occurs precisely at the border links, and then those are the corresponding terms that (simultaneously) will approach the minimum value of $-1$. Then, in this case, we find that the minimum value for the coupling that allows synchronization is $\sigma_c=\frac{N}{4}$.

For the sake of completeness we have also considered 1-dimensional lattices with next-nearest neighbors:

\noindent c) The same geometry as case a) but with 2 additional connections to identical oscillators for each one (see Fig. \ref{fig:regular_lattices}(c)). Now one has $S_{\cal G}=1/2$.
We can also make the same assumption as before and the result is unchanged with respect to a) because the zero phase difference between identical units. Hence we get exactly the  same critical value $\sigma_c = 1/2$.

\noindent d) This corresponds to add two next nearest neighbors to case b) (see Fig. \ref{fig:regular_lattices}(d)) and $S_{\cal G}=1-3/N $. Analogously to case b) we obtain summing the equations for nodes $1, \cdots, \frac{N}{2} $:

\begin{eqnarray}
\label{1d_1sim}
\lefteqn{\dot{\varphi}_1 + \cdots + \dot{\varphi}_{\frac{N}{2}}= } \nonumber \\
&& \frac{N}{2} + \sigma \sin ({\varphi}_{\frac{N}{2} + 1} - {\varphi}_{\frac{N}{2} - 1}) +  \nonumber \\
&& \sigma \sin ({\varphi}_{\frac{N}{2} + 1} - {\varphi}_{\frac{N}{2}}) +  \sigma \sin ({\varphi}_{\frac{N}{2} + 2} - {\varphi}_{\frac{N}{2}}) + \nonumber \\
&& \sigma \sin (\varphi_{N} - \varphi_1) +  \sigma \sin (\varphi_{N - 1} - \varphi_1) + \nonumber \\
&& \sigma \sin (\varphi_{N} - \varphi_2).
\end{eqnarray}
As before, we can assume that the difference between these phases maximizes the modulus of the sine terms. Then in this case we have, $\sigma_c=\frac{N/2}{6}=N/12$.

\begin{figure}
\includegraphics[width=0.5\textwidth]{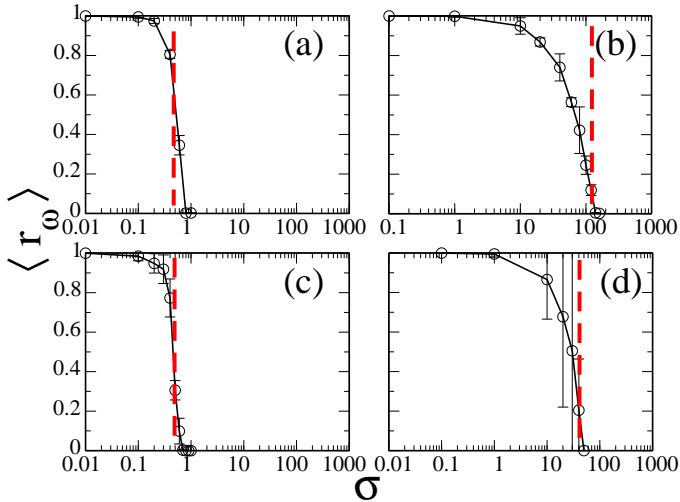}
\caption{\label{fig:example} Synchronization in 1-d regular networks ($N = 500$) for different degrees and network similarities. Labels correspond to those of Fig. \ref{fig:regular_lattices}. Averaged frequency dispersion ($\langle r_{\omega} \rangle$) as a function of the coupling $\sigma$. Vertical dashed lines indicate our analytical estimate for the minimum coupling strength needed for synchronization $\sigma_c$ (0.5 in cases (a) and (c), 125 in case (b) and 41.6 in case (d)).}
\end{figure}

In general, we can say that for this simple and symmetric structures, the predicted values agree very well with those obtained in the simulations. For cases a) and c) the only assumption is that white and black clusters have common effective frequencies and phases, although different among them. However, in cases b) and d) we rely on the fact that the maximum phase difference between any two neighboring nodes is produced at the border between black and white clusters. This is what we have also observed in the simulations (see Fig. \ref{fig:example}).

\section{Generalization to random networks}
\label{sec5}

Finally, we present a possible extension of these results to the general case of random networks.
Let us consider a network with a very low similarity, so we can assume that a unit with natural frequency $\omega_i$
is surrounded by $N_i(\omega_i)>0$ neighbors with the same frequency and by $N_i-N_i(\omega_i)$ nodes of opposite sign.
We can assume as for case a) in the previous section, now it is an approximation not as exact as before,  that all positive units have the same frequency and $\dot{\varphi}_+$ and phase
${\varphi}_+$, and the same assumption for the negative ones. Under this hypothesis we have :
\begin{equation}
\label{eq:approx}
\dot{\varphi}_i=\omega_i+(N_i-N_i(\omega_i)) \sigma \sin (\varphi_{-\omega_i} - \varphi_{\omega_i})
\end{equation}
and then the minimum value of the coupling to compensate the frequency $|\omega_i|=1$ is $1/(N_i-N_i(\omega_i))$. This has to happen for all the nodes
simultaneously in the network and, hence, we will have to look for the global maximum:
\begin{equation}
\label{eq:sigma_c_low}
\sigma_c=\max_i \frac{1}{N_i-N_i(\omega_i)}
\end{equation}

On the contrary, for high similarity networks we can proceed in a similar way as in cases b) and d) before.
Take a group, A, of connected oscillators with the same natural frequency $\omega_A$ surrounded by oscillators with frequency $-\omega_A$, and sum up all their equations in the synchronized state
\begin{equation}
\label{eq:approx_high_sim}
0=N_A\cdot\omega_A+\sigma \sum_{i\in A,j\not\in A} a_{ij} \sin (\varphi_{j} - \varphi_i).
\end{equation}
Now we can make the approximate assumption that the phase difference that maximizes the modulus of the sine terms again
is at the borders between groups of different sign. Then the minimum value of $\sigma$ that can fulfill eq. (\ref{eq:approx_high_sim}) is
$N_A/L_A$, where $L_A$ is the number of links pointing out of the group $A$. Again, since this has to be verified for any group, the critical
value of the coupling will be
\begin{equation}
\label{eq:sigma_c_high}
\sigma_c=\max_A \frac{N_A}{L_A}
\end{equation}
which contains our previous result for isolated nodes (\ref{eq:sigma_c_low}) as a particular case. In this way, we can take eq. (\ref{eq:sigma_c_high})
as a result valid for any configuration, no matter how low or high is the similarity. We have tested this approach on scale-free and regular random networks finding good agreement between this prediction and numerical results (see Fig \ref{fig:exp_RR+SF}). Our result is exact only for a particular case of symmetric organization. However, notice that simple ideas providing results on regular networks are only rarely so easy to generalize to any kind of network topology.
Furthermore, this analysis coincides with the qualitative explanation at the end of Sect. III. We discussed there that the global synchronization for networks of high frequency similarity was burdened by the size of large regions of identical frequency, which explain why they are harder to synchronize than networks
of lower frequency similarity.\\
\section{Conclusions}
\label{sec6}
In this paper we have studied synchronization in the population of Kuramoto oscillators with a bipolar distribution of natural frequencies. Such scenario can, for example, be relevant for special cases when the network is optimized with respect to local or global synchronization \cite{brede07, brede08}. It can also be seen as a proxy model for power grid networks where there are two complementary populations of nodes either generating or consuming energy, and the final goal is to achieve a complete synchronized state where electricity is distributed in a well defined global frequency \cite{fnp07}.

First, we have analyzed the synchronization on random networks under different conditions,  controlling the level of frequency similarity. We introduce frequency similarity of a network as a measure of the frequency similarity of neighboring nodes with respect to the frequency they have. We find a strong impact of the topological distribution of frequencies on the synchronization process. When the frequency similarity is low, the effect of network topology is only marginal.  When increasing the similarity, the role of network topology becomes stronger, e.g. synchronization in scale-free networks  becomes easier than in regular random networks due to the influence of hubs.

Furthermore, we have gone beyond the results in \cite{brede07} and \cite{brede08} by calculating exactly the minimum value of coupling strength that is necessary for global synchronization in simple regular networks. Generalizing this particular case, we obtain a simple method to estimate the synchronization threshold for arbitrary network topologies. Altogether, it is sufficient to count  the size $N_A$ of interconnected clusters of nodes, which have the same natural frequency $\omega_A$ and the number of links $L_A$ connecting these clusters with the nodes of opposite frequency. Then, the maximal ratio $N_A/L_A$ calculated over all clusters present in the network gives us an estimate for the coupling strength sufficient for global synchronization. This general result computed in terms of the topological distribution of opposite frequencies provides an accurate estimation of a dynamical property of the complete network.
\begin{acknowledgments}
L.B. and S.L. gratefully acknowledge partial financial support by the EU project IRRIIS No. 027568. A.D.-G. thanks Ministerio de Educaci\'{o}n y Ciencia (FIS-2006-13321 and PR2008-0114) ) and Generalitat de Catalunya (2005SGR00889). We furthermore thank A. Arenas and D. Helbing for comments and suggestions.
\end{acknowledgments}
\bibliographystyle{apsrev}

\begin{thebibliography}{30}
\expandafter\ifx\csname natexlab\endcsname\relax\def\natexlab#1{#1}\fi
\expandafter\ifx\csname bibnamefont\endcsname\relax
  \def\bibnamefont#1{#1}\fi
\expandafter\ifx\csname bibfnamefont\endcsname\relax
  \def\bibfnamefont#1{#1}\fi
\expandafter\ifx\csname citenamefont\endcsname\relax
  \def\citenamefont#1{#1}\fi
\expandafter\ifx\csname url\endcsname\relax
  \def\url#1{\texttt{#1}}\fi
\expandafter\ifx\csname urlprefix\endcsname\relax\def\urlprefix{URL }\fi
\providecommand{\bibinfo}[2]{#2}
\providecommand{\eprint}[2][]{\url{#2}}

\bibitem[{\citenamefont{{Strogatz}}(2001)}]{s01}
\bibinfo{author}{\bibfnamefont{S.~H.} \bibnamefont{{Strogatz}}},
  \bibinfo{journal}{Nature (London)} \textbf{\bibinfo{volume}{410}},
  \bibinfo{pages}{268} (\bibinfo{year}{2001}).

\bibitem[{\citenamefont{{Albert} and {Barab{\'a}si}}(2002)}]{ab02}
\bibinfo{author}{\bibfnamefont{R.}~\bibnamefont{{Albert}}} \bibnamefont{and}
  \bibinfo{author}{\bibfnamefont{A.-L.} \bibnamefont{{Barab{\'a}si}}},
  \bibinfo{journal}{Rev. Mod. Phys.} \textbf{\bibinfo{volume}{74}},
  \bibinfo{pages}{47} (\bibinfo{year}{2002}).

\bibitem[{\citenamefont{Dorogovtsev and Mendes}(2002)}]{dm02}
\bibinfo{author}{\bibfnamefont{S.}~\bibnamefont{Dorogovtsev}} \bibnamefont{and}
  \bibinfo{author}{\bibfnamefont{J.~F.~F.} \bibnamefont{Mendes}},
  \bibinfo{journal}{Adv. Phys.} \textbf{\bibinfo{volume}{51}},
  \bibinfo{pages}{1079} (\bibinfo{year}{2002}).

\bibitem[{\citenamefont{{Newman}}(2003)}]{n03a}
\bibinfo{author}{\bibfnamefont{M.~E.~J.} \bibnamefont{{Newman}}},
  \bibinfo{journal}{SIAM Rev.} \textbf{\bibinfo{volume}{45}},
  \bibinfo{pages}{167} (\bibinfo{year}{2003}).

\bibitem[{\citenamefont{{Boccaletti} et~al.}(2006)\citenamefont{{Boccaletti},
  {Latora}, {Moreno}, {Chavez}, and {Hwang}}}]{blmch06}
\bibinfo{author}{\bibfnamefont{S.}~\bibnamefont{{Boccaletti}}},
  \bibinfo{author}{\bibfnamefont{V.}~\bibnamefont{{Latora}}},
  \bibinfo{author}{\bibfnamefont{Y.}~\bibnamefont{{Moreno}}},
  \bibinfo{author}{\bibfnamefont{M.}~\bibnamefont{{Chavez}}}, \bibnamefont{and}
  \bibinfo{author}{\bibfnamefont{D.-U.} \bibnamefont{{Hwang}}},
  \bibinfo{journal}{Phys. Rep.} \textbf{\bibinfo{volume}{424}},
  \bibinfo{pages}{175} (\bibinfo{year}{2006}).

\bibitem[{\citenamefont{da~Fontoura~Costa
  et~al.}(2007)\citenamefont{da~Fontoura~Costa, Rodrigues, Travieso, and
  Boas}}]{crtbv07}
\bibinfo{author}{\bibfnamefont{L.}~\bibnamefont{da~Fontoura~Costa}},
  \bibinfo{author}{\bibfnamefont{F.~A.} \bibnamefont{Rodrigues}},
  \bibinfo{author}{\bibfnamefont{G.}~\bibnamefont{Travieso}}, \bibnamefont{and}
  \bibinfo{author}{\bibfnamefont{P.~R.~V.} \bibnamefont{Boas}},
  \bibinfo{journal}{Adv. Phys.} \textbf{\bibinfo{volume}{56}},
  \bibinfo{pages}{167} (\bibinfo{year}{2007}).

\bibitem[{\citenamefont{{Newman}}(2002)}]{n02}
\bibinfo{author}{\bibfnamefont{M.~E.~J.} \bibnamefont{{Newman}}},
  \bibinfo{journal}{Phys. Rev. Lett.} \textbf{\bibinfo{volume}{89}},
  \bibinfo{pages}{208701} (\bibinfo{year}{2002}).

\bibitem[{\citenamefont{Serrano et~al.}(2007)\citenamefont{Serrano,
  Bogu{\~n}{\'a}, Pastor-Satorras, and Vespignani}}]{sbpv07}
\bibinfo{author}{\bibfnamefont{M.~A.} \bibnamefont{Serrano}},
  \bibinfo{author}{\bibfnamefont{M.}~\bibnamefont{Bogu{\~n}{\'a}}},
  \bibinfo{author}{\bibfnamefont{R.}~\bibnamefont{Pastor-Satorras}},
  \bibnamefont{and}
  \bibinfo{author}{\bibfnamefont{A.}~\bibnamefont{Vespignani}}, in
  \emph{\bibinfo{booktitle}{Structure and Dynamics of Complex Networks, From
  Information Technology to Finance and Natural Science}}
  (\bibinfo{publisher}{World Scientific}, \bibinfo{address}{Singapore},
  \bibinfo{year}{2007}), pp. \bibinfo{pages}{35--66}.

\bibitem[{\citenamefont{Zhou and Mondragon}(2004)}]{rich-club}
\bibinfo{author}{\bibfnamefont{S.}~\bibnamefont{Zhou}} \bibnamefont{and}
  \bibinfo{author}{\bibfnamefont{R.~J.} \bibnamefont{Mondragon}},
  \bibinfo{journal}{Communications Letters, IEEE} \textbf{\bibinfo{volume}{8}},
  \bibinfo{pages}{180} (\bibinfo{year}{2004}).

\bibitem[{\citenamefont{Girvan and Newman}(2002)}]{gn02}
\bibinfo{author}{\bibfnamefont{M.}~\bibnamefont{Girvan}} \bibnamefont{and}
  \bibinfo{author}{\bibfnamefont{M.~E.~J.} \bibnamefont{Newman}},
  \bibinfo{journal}{Proc. Natl. Acad. Sci. USA} \textbf{\bibinfo{volume}{99}},
  \bibinfo{pages}{7821} (\bibinfo{year}{2002}).

\bibitem[{\citenamefont{{Danon} et~al.}(2005)\citenamefont{{Danon},
  {D{\'{\i}}az-Guilera}, {Duch}, and {Arenas}}}]{ddda05}
\bibinfo{author}{\bibfnamefont{L.}~\bibnamefont{{Danon}}},
  \bibinfo{author}{\bibfnamefont{A.}~\bibnamefont{{D{\'{\i}}az-Guilera}}},
  \bibinfo{author}{\bibfnamefont{J.}~\bibnamefont{{Duch}}}, \bibnamefont{and}
  \bibinfo{author}{\bibfnamefont{A.}~\bibnamefont{{Arenas}}},
  \bibinfo{journal}{J. Stat. Mech.} \textbf{\bibinfo{volume}{9}},
  \bibinfo{pages}{8} (\bibinfo{year}{2005}).

\bibitem[{\citenamefont{Barrat et~al.}(2008)\citenamefont{Barrat, Barthelemy,
  and Vespignani}}]{indiana_book}
\bibinfo{author}{\bibfnamefont{A.}~\bibnamefont{Barrat}},
  \bibinfo{author}{\bibfnamefont{M.}~\bibnamefont{Barthelemy}},
  \bibnamefont{and}
  \bibinfo{author}{\bibfnamefont{A.}~\bibnamefont{Vespignani}},
  \emph{\bibinfo{title}{Dynamical processes on complex networks}}
  (\bibinfo{publisher}{Cambridge University Press}, \bibinfo{year}{2008}).

\bibitem[{\citenamefont{Strogatz}(2003)}]{strogatz03}
\bibinfo{author}{\bibfnamefont{S.~H.} \bibnamefont{Strogatz}},
  \emph{\bibinfo{title}{Sync: The Emerging Science of Spontaneous Order}}
  (\bibinfo{publisher}{Hyperion}, \bibinfo{address}{New York, NY, USA},
  \bibinfo{year}{2003}).

\bibitem[{\citenamefont{Kuramoto}(1984)}]{Kuramoto84}
\bibinfo{author}{\bibfnamefont{Y.}~\bibnamefont{Kuramoto}},
  \emph{\bibinfo{title}{Chemical oscillations, waves, and turbulence}}
  (\bibinfo{publisher}{Springer-Verlag}, \bibinfo{address}{New York, NY, USA},
  \bibinfo{year}{1984}).

\bibitem[{\citenamefont{{Acebr{\'o}n} et~al.}(2005)\citenamefont{{Acebr{\'o}n},
  {Bonilla}, {P{\'e}rez-Vicente}, {Ritort}, and {Spigler}}}]{abprs05}
\bibinfo{author}{\bibfnamefont{J.~A.} \bibnamefont{{Acebr{\'o}n}}},
  \bibinfo{author}{\bibfnamefont{L.~L.} \bibnamefont{{Bonilla}}},
  \bibinfo{author}{\bibfnamefont{C.~J.} \bibnamefont{{P{\'e}rez-Vicente}}},
  \bibinfo{author}{\bibfnamefont{F.}~\bibnamefont{{Ritort}}}, \bibnamefont{and}
  \bibinfo{author}{\bibfnamefont{R.}~\bibnamefont{{Spigler}}},
  \bibinfo{journal}{Rev. Mod. Phys.} \textbf{\bibinfo{volume}{77}},
  \bibinfo{pages}{137} (\bibinfo{year}{2005}).

\bibitem[{\citenamefont{Arenas et~al.}(2008)\citenamefont{Arenas,
  D{\'i}az-Guilera, Kurths, Moreno, and Zhou}}]{adkmz08}
\bibinfo{author}{\bibfnamefont{A.}~\bibnamefont{Arenas}},
  \bibinfo{author}{\bibfnamefont{A.}~\bibnamefont{D{\'i}az-Guilera}},
  \bibinfo{author}{\bibfnamefont{J.}~\bibnamefont{Kurths}},
  \bibinfo{author}{\bibfnamefont{Y.}~\bibnamefont{Moreno}}, \bibnamefont{and}
  \bibinfo{author}{\bibfnamefont{C.}~\bibnamefont{Zhou}},
  \bibinfo{journal}{Phys. Rep.} \textbf{\bibinfo{volume}{469}},
  \bibinfo{pages}{93} (\bibinfo{year}{2008}).

\bibitem[{\citenamefont{Brede}(2008)}]{brede07}
\bibinfo{author}{\bibfnamefont{M.}~\bibnamefont{Brede}},
  \bibinfo{journal}{Phys. Lett. A} \textbf{\bibinfo{volume}{372}},
  \bibinfo{pages}{2618} (\bibinfo{year}{2008}).

\bibitem[{\citenamefont{{Gleiser} and {Zanette}}(2006)}]{gz06}
\bibinfo{author}{\bibfnamefont{P.~M.} \bibnamefont{{Gleiser}}}
  \bibnamefont{and} \bibinfo{author}{\bibfnamefont{D.~H.}
  \bibnamefont{{Zanette}}}, \bibinfo{journal}{Europ. Phys. J. B}
  \textbf{\bibinfo{volume}{53}}, \bibinfo{pages}{233} (\bibinfo{year}{2006}).

\bibitem[{\citenamefont{Martens et~al.}(2009)\citenamefont{Martens, Barreto,
  Strogatz, Ott, So, and Antonsen}}]{Martens}
\bibinfo{author}{\bibfnamefont{E.A.}~\bibnamefont{Martens}},
  \bibinfo{author}{\bibfnamefont{E.}~\bibnamefont{Barreto}},
  \bibinfo{author}{\bibfnamefont{S.H.}~\bibnamefont{Strogatz}},
  \bibinfo{author}{\bibfnamefont{E.}~\bibnamefont{Ott}},
  \bibinfo{author}{\bibfnamefont{P.}~\bibnamefont{So}}, \bibnamefont{and}
  \bibinfo{author}{\bibfnamefont{T.M.}~\bibnamefont{Antonsen}},
  \bibinfo{journal}{Phys. Rev. E} \textbf{\bibinfo{volume}{79}},
  \bibinfo{pages}{026204} (\bibinfo{year}{2009}).

\bibitem[{\citenamefont{Montbri{\'o} et~al.}(2006)\citenamefont{Montbri{\'o},
  Paz{\'o}, and Schmidt}}]{Montbrio}
\bibinfo{author}{\bibfnamefont{E.}~\bibnamefont{Montbri{\'o}}},
  \bibinfo{author}{\bibfnamefont{D.}~\bibnamefont{Paz{\'o}}}, \bibnamefont{and}
  \bibinfo{author}{\bibfnamefont{J.}~\bibnamefont{Schmidt}},
  \bibinfo{journal}{Phys. Rev. E} \textbf{\bibinfo{volume}{74}}
    \bibinfo{pages}{056201} (\bibinfo{year}{2006}).

\bibitem[{\citenamefont{Bonilla et~al.}(1998)\citenamefont{Bonilla,
  P\'{e}rez~Vicente, and Spigler}}]{bps98}
\bibinfo{author}{\bibfnamefont{L.~L.} \bibnamefont{Bonilla}},
  \bibinfo{author}{\bibfnamefont{C.~J.} \bibnamefont{P\'{e}rez~Vicente}},
  \bibnamefont{and} \bibinfo{author}{\bibfnamefont{R.}~\bibnamefont{Spigler}},
  \bibinfo{journal}{Physica D} \textbf{\bibinfo{volume}{113}},
  \bibinfo{pages}{79} (\bibinfo{year}{1998}).

\bibitem[{\citenamefont{{Filatrella} et~al.}(2008)\citenamefont{{Filatrella},
  {Nielsen}, and {Pedersen}}}]{fnp07}
\bibinfo{author}{\bibfnamefont{G.}~\bibnamefont{{Filatrella}}},
  \bibinfo{author}{\bibfnamefont{A.~H.} \bibnamefont{{Nielsen}}},
  \bibnamefont{and} \bibinfo{author}{\bibfnamefont{N.~F.}
  \bibnamefont{{Pedersen}}}, \bibinfo{journal}{Europ. Phys. J. B}
  \textbf{\bibinfo{volume}{61}}, \bibinfo{pages}{485} (\bibinfo{year}{2008}).

\bibitem[{\citenamefont{Acebr{\'o}n and Spigler}(1998)}]{Acebron}
\bibinfo{author}{\bibfnamefont{J.A.}~\bibnamefont{Acebr{\'o}n}} \bibnamefont{and}
  \bibinfo{author}{\bibfnamefont{R.}~\bibnamefont{Spigler}},
  \bibinfo{journal}{Phys. Rev. Lett.} \textbf{\bibinfo{volume}{81}}
      \bibinfo{pages}{2229} (\bibinfo{year}{1998}).

\bibitem[{\citenamefont{Tanaka and Lichtenberg}(1997)}]{Tanaka}
\bibinfo{author}{\bibfnamefont{H.A.}~\bibnamefont{Tanaka}} \bibnamefont{and}
\bibinfo{author}{\bibfnamefont{A.J.}~\bibnamefont{Lichtenberg}},
\bibinfo{author}{\bibfnamefont{S.}~\bibnamefont{Oishi}},
  \bibinfo{journal}{Phys. Rev. Lett.} \textbf{\bibinfo{volume}{78}},
  \bibinfo{pages}{2104 - 2107} (\bibinfo{year}{1997}).

\bibitem[{\citenamefont{Barab\'asi and Albert}(1999)}]{ba99}
\bibinfo{author}{\bibfnamefont{A.~L.} \bibnamefont{Barab\'asi}}
  \bibnamefont{and} \bibinfo{author}{\bibfnamefont{R.}~\bibnamefont{Albert}},
  \bibinfo{journal}{Science} \textbf{\bibinfo{volume}{286}},
  \bibinfo{pages}{509} (\bibinfo{year}{1999}).

\bibitem[{\citenamefont{Kim and Vu}(2003)}]{kim_2003}
\bibinfo{author}{\bibfnamefont{J.~H.} \bibnamefont{Kim}} \bibnamefont{and}
  \bibinfo{author}{\bibfnamefont{V.~H.} \bibnamefont{Vu}}, in
  \emph{\bibinfo{booktitle}{Proceedings of the thirty-fifth ACM symposium on
  Theory of computing}} (\bibinfo{publisher}{ACM Press}, \bibinfo{year}{2003}),
  pp. \bibinfo{pages}{213--222}, ISBN \bibinfo{isbn}{1-58113-674-9}.

\bibitem[{\citenamefont{{Sendi{\~n}a-Nadal}
  et~al.}(2008)\citenamefont{{Sendi{\~n}a-Nadal}, Buldu, Leyva, and
  Boccaletti}}]{Nadal}
\bibinfo{author}{\bibfnamefont{I.}~\bibnamefont{{Sendi{\~n}a-Nadal}}},
  \bibinfo{author}{\bibfnamefont{J.}~\bibnamefont{Buldu}},
  \bibinfo{author}{\bibfnamefont{I.}~\bibnamefont{Leyva}}, \bibnamefont{and}
  \bibinfo{author}{\bibfnamefont{S.}~\bibnamefont{Boccaletti}},
  \bibinfo{journal}{PLoS ONE} \textbf{\bibinfo{volume}{3}}
  (\bibinfo{year}{2008}).

\bibitem[{\citenamefont{{G\'omez-Garde\~nes}
  et~al.}(2007)\citenamefont{{G\'omez-Garde\~nes}, {Moreno}, and
  {Arenas}}}]{gma07a}
\bibinfo{author}{\bibfnamefont{J.}~\bibnamefont{{G\'omez-Garde\~nes}}},
  \bibinfo{author}{\bibfnamefont{Y.}~\bibnamefont{{Moreno}}}, \bibnamefont{and}
  \bibinfo{author}{\bibfnamefont{A.}~\bibnamefont{{Arenas}}},
  \bibinfo{journal}{Phys. Rev. Lett.} \textbf{\bibinfo{volume}{98}},
  \bibinfo{pages}{034101} (\bibinfo{year}{2007}).

\bibitem[{\citenamefont{Brede}(2008)}]{brede08}
\bibinfo{author}{\bibfnamefont{M.}~\bibnamefont{Brede}},
  \bibinfo{journal}{Europ. Phys. J. B} \textbf{\bibinfo{volume}{62}},
  \bibinfo{pages}{87} (\bibinfo{year}{2008}).

\bibitem[{\citenamefont{{Arenas} et~al.}(2006)\citenamefont{{Arenas},
  {D{\'{\i}}az-Guilera}, and {P{\'e}rez-Vicente}}}]{adp06a}
\bibinfo{author}{\bibfnamefont{A.}~\bibnamefont{{Arenas}}},
  \bibinfo{author}{\bibfnamefont{A.}~\bibnamefont{{D{\'{\i}}az-Guilera}}},
  \bibnamefont{and} \bibinfo{author}{\bibfnamefont{C.~J.}
  \bibnamefont{{P{\'e}rez-Vicente}}}, \bibinfo{journal}{Phys. Rev. Lett.}
  \textbf{\bibinfo{volume}{96}}, \bibinfo{pages}{114102}
  (\bibinfo{year}{2006}).
\end{thebibliography}

\end{document}